\documentclass[table, hidelinks]{preprint}
\usepackage[utf8]{inputenc}
\usepackage[T1]{fontenc}
\usepackage{graphicx}
\usepackage{longtable}
\usepackage{wrapfig}
\usepackage{rotating}
\usepackage[normalem]{ulem}
\usepackage{amsmath}
\usepackage{amssymb}
\usepackage{capt-of}
\usepackage{hyperref}
\usepackage{xcolor}
\usepackage{booktabs}
\usepackage{ebgaramond}
\newbox{\myorcidaffilbox}
\sbox{\myorcidaffilbox}{\large\includegraphics[height=1.7ex]{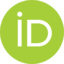}}
\newcommand{\orcidaffil}[1]{%
\href{https://orcid.org/#1}{\usebox{\myorcidaffilbox}}}
\newbox{\myemailbox}
\sbox{\myemailbox}{\large\includegraphics[height=1.7ex]{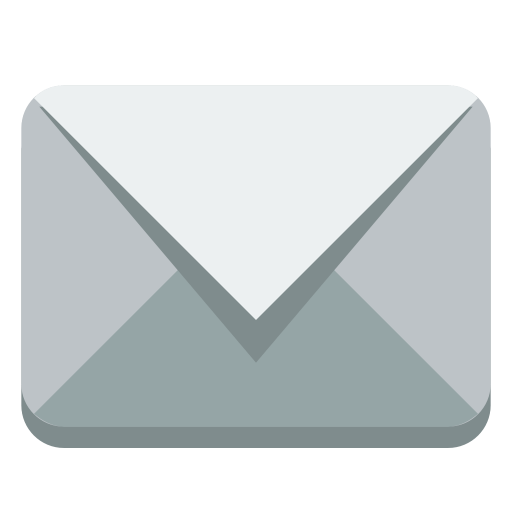}}
\newcommand{\emailicon}[1]{%
\href{mailto:#1?Subject=Re:Digital Innovation in Microenterprises: Current Trends and New Research Avenues}{\usebox{\myemailbox}}}
\usepackage{authblk}
\usepackage{orcidlink}
\author[1]{Juan E. Gómez-Morantes \orcidaffil{0000-0002-8107-4030}\emailicon{je.gomezm@javeriana.edu.co}, PhD}
\author[2]{Andrea Herrera \orcidaffil{0000-0003-2898-7161}, PhD}
\author[3]{Sonia Camacho \orcidaffil{0000-0002-3662-1402}, PhD}
\affil[1]{Departamento de Ingeniería de Sistemas, Pontificia Universidad Javeriana, Bogotá, Colombia}
\affil[2]{Departamento de Ingeniería de Sistemas y Computación, Universidad de los Andes, Bogotá, Colombia}
\affil[3]{Facultad de Administración, Universidad de los Andes, Bogotá, Colombia}
\abstract{The relationship between microenterprises and information and communication technologies (ICTs) has always been troublesome. Because of the rapid pace of modern digital technologies, digital innovation processes are permeating the industries, markets, and social contexts in which microenterprises exist today. However, microenterprises have severe difficulties engaging or performing in these digital contexts and are at risk of being left behind. This paper reviews the literature on ICTs and microenterprises, focusing on the adoption, usage, and impact of ICTs. The results indicate that further research in this field should avoid focusing on individual microenterprises (or samples of independent microenterprises) as the unit of analysis and should favour a systemic approach in which markets, value chains, or microenterprise-intensive sectors are studied. Additionally, theoretical frameworks capable of considering change and the dynamic nature of innovation processes are highlighted as a critical focus area for the field.}
\usepackage[citestyle=authoryear,bibstyle=authoryear,backend=biber,maxbibnames=3,isbn=false,uniquename=minfull,maxcitenames=2] {biblatex}
\AtEveryBibitem{\clearfield{pagetotal}}
\AtEveryBibitem{\clearfield{url}}
\AtEveryBibitem{\clearfield{urldate}}
\date{\today}
\title{Digital Innovation in Microenterprises: Current Trends and New Research Avenues}
\hypersetup{
 pdfauthor={},
 pdftitle={Digital Innovation in Microenterprises: Current Trends and New Research Avenues},
 pdfkeywords={},
 pdfsubject={},
 pdfcreator={},
 pdflang={English}}
\begin{document}

\maketitle

\TPJcopyright{This is an open access paper under the terms of the Creative Commons Attribution-NonCommercial License, which permits use, distribution, and reproduction in any medium, provided the original work is properly cited and is not used for commercial purposes. More information about the license can be found at the following link: \href{https://creativecommons.org/licenses/by-nc/4.0/}{CC-BY-NC License}.}
\section{Introduction}
\label{sec:org3be7048}
Advances in information and communication technologies (ICTs) have placed such technologies at the centre of social functions such as politics, interpersonal relations, education, national security, and environmental conservation. As a result, the digital economy is now commonplace, with businesses and markets migrating to the digital realm at an alarming pace \autocite{Freeman2001,WorldBank2016}. In this context, exploring the relationship between ICTs and business in all its facets is imperative. While the literature on this subject is not scarce, it mostly focuses on large enterprises. However, because of the fundamental differences between microenterprises and large firms, this literature does not offer enough insights into the relationship between ICTs and microenterprises. Even though microenterprises have historically been viewed as scaled-down versions of large firms \autocite{Westhead1996}, this perspective obscures their true nature and results in technologies, processes, and approaches that are watered-down versions of large firms that neglect the inherent characteristics of microenterprises.  

The lack of research in this subject is worrisome because microenterprises serve essential social and economic roles and constitute the basis of the economic pyramid in most countries \autocite{Schreiner2003}, therefore deserving special attention.  

To grasp the full complexity of the relationship between ICTs and microenterprises, this paper reviews our current understanding of ICT innovation processes in microenterprises, covering the adoption, use, and impact of ICTs in this sector. The following section discusses the challenges of tackling the literature on microenterprises and ICTs. Afterwards, this literature is reviewed and discussed in depth. Finally, the paper presents a series of conclusions and recommendations for future research in this field.  
\section{Delimiting the Literature on Digital Innovation in Microenterprises}
\label{sec:orgb12e60d}
ICT and microenterprises have been researched in fields such as management, information systems, economics, and innovation studies \autocite[see for example][]{Burke2009,Qureshil2009,Sandberg2020}. Therefore, a clear delimitation of the literature is necessary for a coherent analysis. The first consideration pertains to the concept of ICT. In this paper, ICTs are defined as technologies---networks of heterogeneous elements (e.g., human, social, institutional, technical) working together to perform a function \autocite{Hughes1986}---that facilitates the capture, storage, processing, or communication of information. While this includes technologies such as notebooks or bulletin boards, this paper focuses on digital technologies (e.g., mobile phones, computers, enterprise software) because of their increasing importance in modern society \autocite{WorldBank2016}. 

The second consideration involves the process of innovation and the limited roles that microenterprises assume in them. Although microenterprises can indeed show extensive uses of technology \autocite{Heeks2008}, where technological elements are part of their value offer, intensive uses of technology (i.e., ICT applied to existing products or processes) are more common in microenterprise sectors. Because of this, and as a complement to previous research on extensive uses of ICT by microenterprises \autocite[see][]{Foster2010}, this research is focused on intensive uses of ICT by microenterprises. 

The next consideration involves the criteria used to classify a firm as a microenterprise. Although terms such as microenterprises or small and mid-size enterprises (SMEs) are used worldwide, their meaning can vary significantly from one region to another \autocite{Kushnir2010}. Perhaps because of this ambiguity, there is a tendency to include these three distinct groups of enterprises---micro, small, and mid-size enterprises---as a single group \autocites(e.g.,)(){Redoli2008}[][]{Kushnir2010}[][]{Golding2008}. These broad classifications, however, overlook the fundamental differences among these groups, thus undermining the applicability of the findings. It is therefore important to start this review with a clear definition of these categories \autocite{Milis2008}. 

Of all the variables used to classify microenterprises, the number of employees is by far the most accepted \autocites{Nichter2009}[][]{Kushnir2010}[][]{Heeks2008}. Furthermore, other factors like turnover or assets are hard to operationalise due to microenterprises' lack of reliable financial records and reluctance to disclose them \autocites{Esselaar2007}[][]{Chew2010}. Therefore, this paper adopts a definition of microenterprises based on employee count. Microenterprises are defined as those with less than ten employees, small enterprises as those with less than 50 employees, and mid-size enterprises as those with less than 200 employees. However, size in terms of employee count can also be misleading. For example, the rise of high-tech start-ups, although small in terms of headcount, are businesses with considerable levels of human, social, and even financial resources \autocite{Quinones2021}, making them difficult to compare with traditional small or micro-enterprises. Because of this, high-tech startups, regardless of their size, are excluded from this review. 

The aforementioned variety of definitions and the tendency to use broad categories like small and medium enterprises (SMEs) or micro, small, and medium enterprises (MSMEs) make navigating the literature difficult. Therefore, microenterprise researchers must choose between a scarce body of literature based on clearly bounded categories (e.g., businesses with less than 20 full-time employees) and a more diverse literature based on broader categories (e.g., MSMEs) but with limited applicability. Given this dilemma, this paper will include research on micro and small enterprises (MSEs)---enterprises with up to 50 employees---to access a broader range of literature without compromising the consistency of the review.  
\section{Methodology}
\label{sec:orgca76b2b}
This review focuses on top journals of four disciplines that traditionally cover the topic of digital innovation processes in MSEs: information systems (IS), innovation studies, ICT for development (ICT4D), and small enterprises. The top IS journals were selected from the first quartile of the Information Science and Library Science category of the InCites Journal Citation Reports. The top innovation studies journals were chosen from the first quartile of the Management of Technology and Innovation category in the SJR SCImago Journal and Country Rank. The top ICT4D journals were selected from the top 10 journals of the ranking developed by \textcite{Heeks2010}. Finally, the top small enterprise journals were selected from a general search with the query ``small enterprises OR small businesses'' in the SJR SCImago Journal and Country Rank, selecting only those in the first or second quartiles of their respective categories. This selection led to a total of 75 journals. Figure 1 shows the process of reviewing these four streams of research to inform our research topic and the choice of primary studies by defining the inclusion and exclusion criteria. Due to the heterogeneous nature of these journals, different queries were used for each discipline\footnote{Sample queries employed:  \\
Information systems journals: \texttt{JN "MIS Quarterly" AND (micro* or small or sme or mse or msb)} \\
Innovation studies journals: \texttt{JN "Academy of Management Review" AND ((micro* OR small OR sme OR mse OR msb) AND ("information technology" OR "information systems" OR ict*))} \\
ICT4D journals: \texttt{JN "Information Technologies \& International Development" AND (micro* or small or sme or mse or msb)} \\
Small enterprises journals: \texttt{JN "Small Enterprise Research" AND ("information technology" OR "information systems" OR ict*)}}, resulting in a list of 859 journal papers.  

\begin{figure}[htbp]
\centering
\includegraphics[width=.9\linewidth]{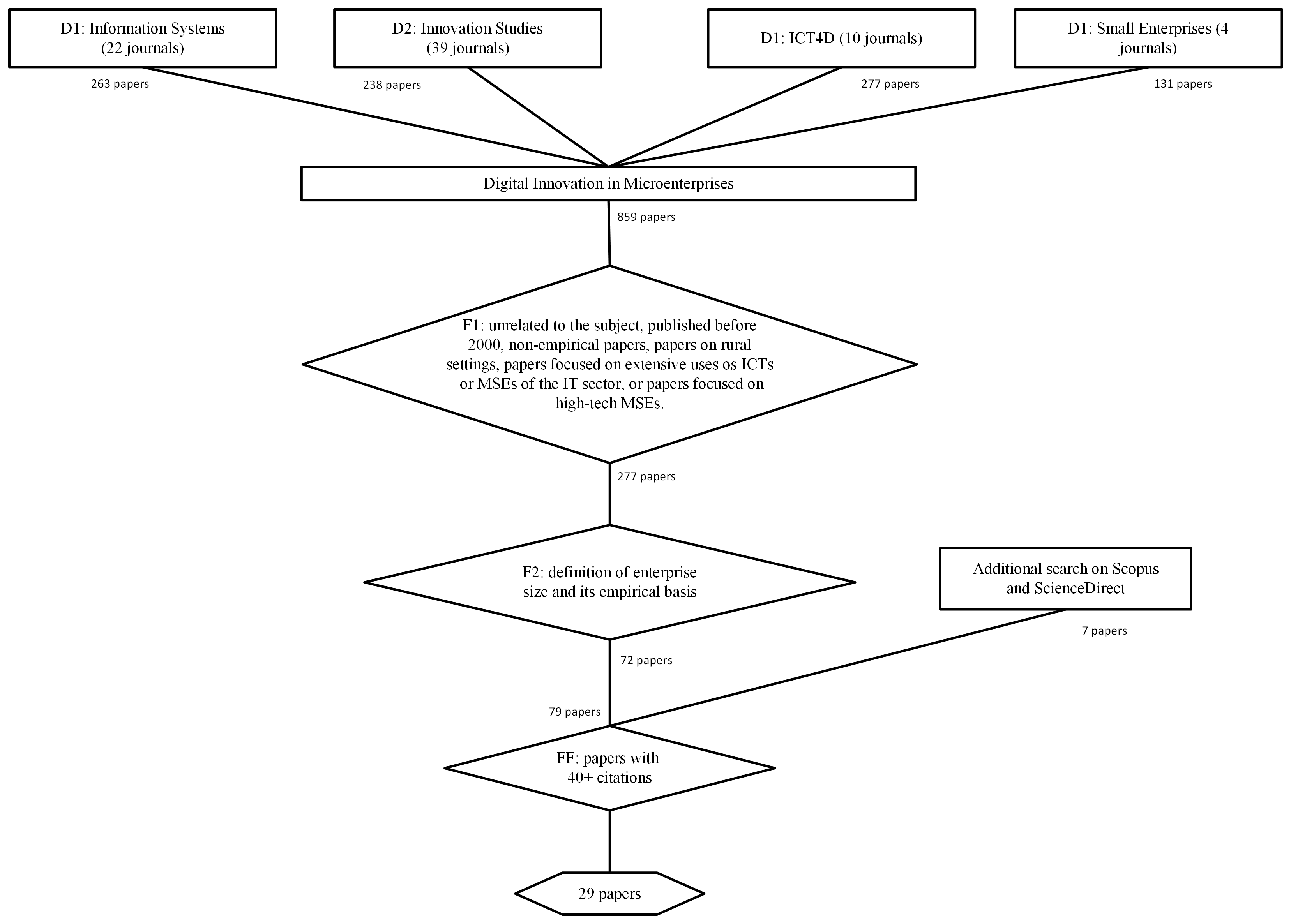}
\caption{\label{fig:filters}Reviewing and revision process}
\end{figure} 

As a first filter (F1 in Figure \ref{fig:filters}), and based on a swift reading of the paper title and abstract, papers with any of the following characteristics were discarded: a) unrelated to the subject of this research, b) published before 2000, c) papers without any empirical evidence, d) papers on rural settings, e) papers focused on extensive uses of ICT or MSEs of the IT sector, or f) papers focused on high-tech MSEs. 

After this initial screening, a total of 632 papers were discarded. Due to inconsistencies in the definition of MSEs, the remaining 227 papers were screened to verify that the definition of enterprise size and their empirical basis matched those in this review (F2 in Figure \ref{fig:filters}). After this second filter, 155 papers were discarded, retaining 72 papers. In addition to these 72 papers, general searches were performed to complement this literature base with relevant papers not published in the journals included in this review. These searches were conducted in Scopus and ScienceDirect. The results of these general searches were subjected to the same screening process described earlier, and seven new papers were identified in Scopus and ScienceDirect, leaving a total of 79 papers. A final filter (FF in Figure \ref{fig:filters}) based on citations was used on this pool of papers, and papers with more than 40 citations in the Google Scholar database were identified as key on the field. These papers are: \autocites{Awiagah2016}[][]{Ahmad2015}{Bengtsson2007}{Chege2020}{Chew2010}{Chew2011}{Chew2015}{Cloete2002}{Donner2004}{Donner2006}{Donner2008a}{Esselaar2007}{Gengatharen2005}{Grimmer2017}{Ilahiane2012}{Jagun2008}{Jones2014}{Kim2017}{Li2018}{MacGregor2005}{Maduku2016}{Maroufkhani2020}{Migiro2005}{Molla2006}{Molony2006}{Moyi2003}{Peltier2009}{Redoli2008}{Wolcott2008}. 
\section{Analysing the literature}
\label{sec:orgc762b59}
Articles were analysed following the approach \textcite{Braun2006} and \textcite{Tsai2021} recommended. The process started by reading the articles and performing a first-order coding based on a three-stage lifecycle model of the relationship between MSEs and ICT: adoption, use, and impact. Such a model is consistent with the innovation adoption stages at organisations \autocite{Damanpour2006,Jasperson2005} and has been used before to analyse ICT impacts \autocite[e.g.,][]{Buchalcevova2012}. The first-order codes were reviewed to identify common themes (e.g., factors internal to MSEs that impact adoption) and grouped accordingly as presented next.  
\subsection{ICT adoption in MSEs}
\label{sec:org60fd76f}
Adoption studies focus on the diffusion of ICTs among MSEs and their adoption processes.  

Adoption is the most studied stage in the literature. Within this category, adoption factors are a dominant theme to the point that some authors claim that it is saturating the literature and that further work on this issue is deemed unnecessary \autocite{Parker2007}. This stream of research investigates the elements that increase or decrease the chances of adoption of a particular and already available ICT. 
\subsubsection{Internal factors: Enterprise size}
\label{sec:org737c5b2}
Some authors have conceptualised enterprise size as an adoption factor. \textcite{Bengtsson2007}, for example, found not only that size is positively correlated with the adoption of advanced Internet applications but also some common adoption factors (e.g., market pressures, top management commitment) among enterprise sizes\footnote{The sizes defined are small (1-19 employees), medium (20-199 employees), and large (more than 200 employees) enterprises.} in Sweden, with entrepreneurial support as a distinctive factor for small enterprises. Other authors exploring enterprise size as a factor in adoption include \textcite{Redoli2008}, who propose a stage model for introducing ICT in SMEs tested in Spain. This model found that smaller enterprises were usually confined to earlier stages of the model. 

A first interpretation of these results is that smaller enterprises are indeed inherently limited in the adoption of ICTs, thus explaining their low adoption rates. However, more developed conceptualisations of the adoption process offer an alternative interpretation. For example, \textcite{Molla2006} conceptualise adoption as an iterative process with multiple sources of influence and feedback points. According to this conceptualisation, adoption is not a one-off process but a sequential process in which enterprises adopt an ICT, react to it, and decide between dropping it or diving further into it. The iterative nature of this process could indicate that MSEs show lower adoption rates than larger enterprises because they are typically younger and have gone through fewer adoption iterations. In addition, size seems to play a role in terms of exposure to what can be described as overwhelming markets in which customer demands (e.g., purchase orders, information, or support requests) far surpass the installed capacity of the business. For example, in their study of microenterprises in the UK, Jones et al. describe the case of a sole-proprietor microenterprise concerned about further ICT adoption generating a demand that ``could not be serviced due to insufficient resources and stock'' \autocite*[, p. 295]{Jones2014}.

Thus, size affects ICT adoption in MSEs due to enterprise youth and an adoption ceiling, where smaller businesses have fewer iterations and reach a point of diminishing returns faster. Since ICT adoption is an iterative and cumulative process \autocite{Molla2006}, this can partially explain the size-related findings of \textcite{Bengtsson2007} and \textcite{Redoli2008}. Meanwhile, the adoption ceiling refers to the point of diminishing returns of the ICT adoption process (i.e., the point at which further adoption iterations will not bring positive outcomes or could even prove harmful to the enterprise). Because of their size and inherent limitations, MSEs seem to reach this adoption ceiling faster than larger organisations, making adopting more advanced or complex ICT pointless. This adoption ceiling, however, is not static; it will grow alongside the enterprise. 
\subsubsection{Internal factors: Enterprise resources}
\label{sec:orge144005}
Authors have analysed internal factors (i.e., the enterprise's or its owner's characteristics) affecting ICT adoption. Although the lack of financial resources is often cited as limiting ICT adoption among MSEs, the evidence is mixed. For example, \textcite{Migiro2005} found costs to be a significant adoption barrier in their study of South African MSEs in the tourism industry. Likewise, \textcite{Jones2014} performed a longitudinal study in 10 microenterprises in the UK and found costs to be a significant variable in ICT adoption. However, other studies have found that financial resources are not that important. \textcite{Peltier2009}, for example, found no correlation between switching costs and CRM adoption in a study involving 386 MSEs in the US. Similarly, \textcite{Moyi2003} studied manufacturing MSEs in Kenya and found that MSEs are willing to pay for reliable information sources, suggesting that they will also be willing to pay for the ICT needed to process, manage, and deliver this information. Despite the literature's mixed results, it is undeniable that MSEs generally adopt more affordable ICT. Mobile phones, for example, have been warmly welcomed in MSEs in developing countries \autocites{Molony2006}[][]{Jagun2008}[][]{Ilahiane2012}[][]{Esselaar2007}[][]{Donner2008a}[][]{Donner2006}[][]{Donner2004}[][]{Chew2010}. Although the success of mobile phones can be attributed to other factors like personal uses given to mobile phones \autocite{Donner2004}, the relatively low cost of devices and airtime is a significant part of this success. 
\subsubsection{Internal factors: Short-term focus}
\label{sec:org178eed9}
The inconsistent results on cost as a factor of adoption could be seen as a matter of return on investment rather than cost. Due to their constraints and vulnerabilities, MSEs are often driven by a short-term mindset \autocite{Jones2014}. This mindset implies that MSEs will be reluctant to invest in a technology that will not result in short-term benefits. Therefore, rather than concluding that MSEs do not have the financial resources needed to adopt ICT, the literature can be better understood as a sign of the preference of MSEs for ICT that a) are more affordable and b) yield short-term returns.  

This short-term focus has been cited as one of the main barriers to adoption in MSEs, implying that a mindset change towards medium- and long-term planning in MSEs is needed to facilitate ICT adoption \autocite{Jones2014}. However, this kind of recommendation forgets that the short-term focus of MSEs is not a choice but an inherent limitation of MSEs and that ``one cannot just wish them away'' \autocite{Molla2006}. It can be concluded, then, that the influence of cost as an adoption factor is rooted in the absence of affordable ICTs compatible with the short-term mindset of MSEs. 
\subsubsection{Internal factors: Others}
\label{sec:orgf61f6b2}
Enterprise and managerial characteristics such as ICT knowledge, risk orientation, top management support, organisational orientation to change and information seeking, and internal skills can be drivers of ICT adoption among MSEs \autocites{Maduku2016}[][]{Molla2006}[][]{Peltier2009}. Internal characteristics such as lack of knowledge or lack of time act, in turn, as internal barriers to ICT adoption \autocites{Jones2014}[][]{MacGregor2005}[][]{Migiro2005}.  
\subsubsection{External factors}
\label{sec:orgf317d51}
In addition to internal aspects, past research has analysed external factors (i.e., access to new markets, competitive pressure, customer pressure, increased contact with customers, market uncertainty, market hostility, provider flexibility, mimetic pressure, government support, enabling conditions, and regulation) \autocites{Awiagah2016}[][]{Jones2014}[][]{Migiro2005}[][]{Molla2006}[][]{Peltier2009}[][]{Ahmad2015} and ICT characteristics (i.e.,  ICT infrastructure, technical complexity, perceived credibility, perceived compatibility with MSEs or their customers, trialability, focus, number of users, resources) \autocites{Awiagah2016}[][]{Gengatharen2005}[][]{Jones2014}[][]{MacGregor2005}[][]{Molla2006}[][]{Peltier2009}. Factors related to the company's supply chain might also act as barriers to ICT adoption, such as low levels of adoption in clients and suppliers \autocites{Awiagah2016}[][]{Cloete2002}.  

It appears to be a wide range of adoption factors. However, it is difficult to generalise any of them or say that they apply to all MSEs. These factors seem primarily dependent on elements such as industry, region, and market conditions. For example, in a sample of 57 MSEs in the tourism sector in South Africa, \textcite{Migiro2005} found an internet adoption rate of 88\%, while \textcite{Chew2010} found an adoption rate of 5.2\% among 231 women-owned MSEs in India. This discrepancy can be explained by the economic sector of the MSEs in question. While the sample of \textcite{Chew2010} was mainly composed of micro retailers with local outreach and some service providers such as physicians, taxicabs, and academic tutoring providers, the sample studied by \textcite{Migiro2005} was composed of tourism and hospitality providers with an international focus. Given that international tourists expect to interact with service providers via the Internet, it makes sense that factors such as ``perceived benefits'' or ``customer pressure'' would be more significant in South African tourism businesses than in Indian micro-retailers. 

There is, however, a factor that presents an interesting reading on the issue of slow ICT adoption by MSEs: network value, which refers to the strong relationship between users and value in connected technologies. As such, the value offered by ICT increases if more users adopt it. This is related to adoption factors like enabling conditions \autocite{Awiagah2016}, perceived benefits \autocites{Migiro2005}[][]{Peltier2009}, critical mass \autocites{Cloete2002}[][]{Gengatharen2005} or compatibility with client's business practices \autocites{Jones2014}[][]{Molla2006}. Since the value of novel ICT is low at the early diffusion stages and builds gradually through the diffusion process \autocite{Rogers2003}, it makes sense for MSEs to be late adopters (i.e., let others create the network value and jump in when it is high enough).  

This network value issue, coupled with the adoption ceiling discussed previously, contradicts the idea that MSEs' non-adoption decisions are based more on intuition and improvisation than reason \autocite{Bengtsson2007}. It seems, then, that the MSEs' decision to delay ICT adoption is not as irrational or near-sighted as sometimes portrayed in the literature. Likewise, the adoption decision is not as unquestionable as some authors suggest \autocite[e.g.,][]{Hairuddin2012}.  

Beyond the MSEs' boundaries, another interesting feature of the adoption process is the actors involved in it. Few studies include customers, suppliers, competitors, and ICT producers as relevant actors. These actors and their influence on the process are usually encapsulated in factors such as customer or provider pressure, market dynamism, and technical knowledge. However, their role and interactions with the MSEs during the adoption process have not been sufficiently explored in the literature. This is surprising because third parties are a frequent feature of adoption processes, especially those dealing with connected technologies. \textcite{Bengtsson2007}, for example, found that Internet adoption for marketing purposes cannot be made in isolation because marketing channels are systemic in nature. Similarly, \textcite{Jagun2008} found that intermediaries actually drove and supported the adoption of mobile phones in the \emph{aso oke}\footnote{The \emph{aso oke} industry produces ceremonial hand-waved clothing.} sector in Nigeria, contradicting the general idea that ICT adoption facilitates disintermediation of industries. Likewise, \textcite{Li2018} found that, as a platform service provider, Alibaba understood that the difficulties faced by many SMEs in launching cross-border e-commerce could be attributed to the management skills of entrepreneurs and organised a series of executive training programmes that offered SMEs the opportunity to learn and share.  
\subsection{ICT use in MSEs}
\label{sec:orgc07771a}
Studies of the use stage focus on the post-adoption relationship between ICTs and MSEs. Regarding use, ICTs can be grouped into two categories: high or low inscription ICTs \autocite{Heeks2002}. Low-inscription ICTs (e.g., word processing applications, the Internet, mobile phones) allow for more diverse usage patterns due to fewer limitations. In contrast, high-inscription ICTs (e.g., ERP systems, e-commerce platforms) do not allow users to deviate from the suppliers' assumptions. Because of their flexibility, low-inscription ICTs have received more attention in the literature, with the mobile phone as a clear frontrunner. 

The uses given to mobile phones by microenterprises have been studied in a variety of locations like India \autocites{Chew2010}[][]{Donner2008a}, Rwanda \autocites{Donner2004}[][]{Donner2006}, Morocco \autocite{Ilahiane2012}, or Tanzania \autocite{Molony2006}. However, like adoption factors, these uses appear highly contingent and context bound. In his work on mobile phone adoption and usage among microenterprises in Rwanda, Donner \autocite*{Donner2004,Donner2006} mentions that because of the low landline access, mobile phones are the only option for most people in the global south to communicate with others. This is contrasted with the situation of MSEs in countries where mobile phones are seen as a complement to landlines. In this situation, landlines are mainly used for business purposes---calls to business contacts---and mobile phones are used for personal purposes---calls to friends and family---\autocites{Chew2010}[][]{Donner2006}, although there is some evidence contradicting this \autocite{Ilahiane2012}.  

Beyond mobile phones, research has been done on the uses given to other ICTs, both with low and high inscriptions. For example, \textcite{Chew2010} also studied the uses given to PCs, email, and Internet connections. Although with far lower adoption rates, those authors found that PCs and email are used more for business than personal uses. However, using the Internet to search for prices or market information---a long-promoted benefit of Internet adoption among MSEs---was found to be insignificant \autocite[cf.][]{Chege2020}. Similarly, \textcite{Molony2006} found that microenterprises use digital communications (mobile phones, text messages, email) more to maintain current social contacts (e.g., increasing the frequency and quality of communications) rather than to create new ones or expand their social network.  

In terms of usage patterns, \textcite{Donner2004} found four perspectives on mobile phone use among microenterprises in Rwanda: a) convenience (i.e., make better use of time and make their life easier), b) intrinsic value (i.e., mobile phone makes them feel more important and use their mobiles for mainly personal reasons), c) indispensability (i.e., microentrepreneurs cite external pressures for their use of the mobile phone and show a heavy component of business uses), and d) productivity (i.e., claiming that their use of mobile phones is either saving them money or increasing their income). As the enterprise grows, it gets closer to the productivity or convenience perspectives. This finding can be attributed to a learning curve in ICT usage, similar to the impact of size as an adoption factor discussed earlier. This implies that as MSEs accumulate knowledge and experience over time (and grow in employee size and complexity), they will show a transition from personal to business uses of mobile phones. In short, the uses of ICT by MSEs are contingent and dynamic. This is consistent with the ideas of \textcite{Molla2006} and suggests that use should also be studied with conceptual frameworks capable of understanding change and evolution over time and interactions with external actors. 
\subsection{ICT impact on MSEs}
\label{sec:orgeff6d86}
Impact studies are focused on the effects of ICT adoption and usage on MSEs or their context. The impacts of ICT adoption and use on MSEs have been widely publicised by practitioners, public policy agents \autocite{WorldBank2016}, and academics \autocites{Cloete2002}[][]{Iacovou1995}. However, there are still open questions about the true impact of ICT on the business world in general \autocites{Brynjolfsson1993}[][]{Brynjolfsson2019} and particularly on MSEs \autocites{Chew2011}[][]{Huaroto2012}. 

There are few but important contributions to empirical research regarding the true impact of ICT on MSEs. \textcite{Huaroto2012}, for example, demonstrates that adopting and using the Internet slightly increases the productivity of Peruvian micro enterprises. However, the mechanisms through which this effect takes place were not explained. Moreover, it is unclear if this increase is sustainable over time if competitors adopt the Internet at the same rate. Similarly, \textcite{Esselaar2007} studied the impact of ICT on microenterprises in 13 African nations and found that ICTs significantly contribute to revenue generation and increased labour productivity. \textcite{Grimmer2017} found that the use of informational resources (e.g., computerised sales system) across Australian small retailers is positively related to their performance (measured as average annual sales turnover). Contrasting these results with recent questions raised about the positive effects of ICT on microenterprises, Esselaar et al. argue that ``the negative impact of ICT investments on business performance reported in the literature can be attributed to the failure to distinguish between the formal and informal sector'' \autocite*[, p. 98]{Esselaar2007}. This is based on the argument that informal MSEs are more profitable and grow faster than formal ones. This means, in essence, that a better distinction between formal and informal microenterprises will present a clearer picture of the true impact of ICT on MSEs' growth and profitability. 

However, in contrast with the previously mentioned sources, other authors are more critical about the impact of ICTs on MSEs. Chew et al. \autocite*{Chew2010,Chew2011} question the benefits that the literature has traditionally attributed to ICT. Their findings indicate that although ICT access is indeed positively related with business growth, other factors like informality are more significant. Additionally, they found that the use of mobile phones for business uses is unrelated and cannot be considered a predictor of business growth. These findings suggest that ICT do contribute to MSEs' growth, but not to the extent usually championed by the ICT industry and taken for granted by some of the literature on this topic. 

Other authors have also explored the impact of ICT on other elements, such as business practices and social capital. \textcite{Donner2006}, for example, suggests mobile phones facilitate the expansion of social networks of microentrepreneurs in Rwanda but do not generate a new network. This makes sense since microentrepreneurs do not call strangers; they call personal or business contacts acquired through referrals or formal introductions. This is supported by \textcite{Molony2006}, who found that digital communications are used more to maintain social and business contacts than to generate them. It is worth noting that microenterprises do not generally use applications intended to create new business connections (e.g., LinkedIn) and tend to focus more on those aimed at maintaining personal contacts (e.g., Facebook, WhatsApp) \autocite{Vatanasakdakul2019}. 

The impact of ICT on broader institutional factors has also been studied. In their work exploring MSEs in Morocco, \textcite{Ilahiane2012} found that although the adoption and use of mobile phones do show some positive outcomes, they are not that significant in the broader poverty alleviation picture. The big issue, they argue, is that poverty responds to a wide range of institutional forces, most outside the scope of markets or market intervention initiatives. This is supported by \textcite{Jagun2008}, who suggest that mobile phones do contribute but are generally unable to solve the structural issues of microenterprise value chains like high costs, speed, and risk exposure. Similarly, \textcite{Donner2008a} argue that mobile phones' impact is closer to amplifying current behaviours and forces than structural changes. On a more positive note, \textcite{Chew2015} make a novel contribution by researching the impact of ICT usage (mainly mobile phones) on mattering---``the perception that people have of how significant they are to others'' \autocite[, p. 524]{Chew2015}---of Indian female microentrepreneurs, finding a positive impact of mobile phone usage on critical aspects of mattering. 

This evidence is a reminder that microenterprises and ICT are embedded in a broader institutional setting with very complex relationships, which should be included in the analysis. This is not only because factors like formality mediate the impacts of ICTs on MSEs but also because adopting a broad enough view is the only way of assessing these impacts. For example, the JIS (just-in-time) production system, with its heavy information and ICT component, is usually seen as highly beneficial to reducing inventory stock and associated costs. Although this might be a fair statement if analysed from an individual-firm perspective, an analysis from a value chain perspective will show that the inventory stock necessities are not reduced nor eliminated but transferred to suppliers further back in the value chain \autocite{Cloete2002}.  Hence, the impact of adopting an ICT will depend on the position occupied by the MSEs within their respective value chains. This again highlights the importance of including broader institutional factors when evaluating the impact of ICT on MSEs. 
\section{Conclusions and future research avenues}
\label{sec:org989c711}
The discussion above shows that the literature covers the adoption, use, and impact stages of ICTs in MSEs well. However, a preference for the adoption stage is evident. This is reasonable because MSEs present lower levels of ICT adoption than other enterprise sizes, which justifies the urgency to understand this phenomenon. Beyond the preference for the adoption stage, this review yields three main conclusions that can be used to guide future research in the field: a) the importance of systemic units of analysis, b) the importance of theories capable of understanding change and evolution, and c) the extent to which research on this field can (or should) be generalised. 

The central gap identified in this literature review is the need for adopting systemic perspectives in the study of digital innovation processes in MSEs since these processes seem to be highly contingent and mediated by a wide range of institutional and contextual elements. As presented in the subsection ``ICT adoption in MSEs'', the adoption behaviour of MSEs is affected by a series of external adoption factors specific to each sector and difficult to generalise. Additionally, ICT adoption does not happen in isolation; MSEs interact with and are influenced by external actors like suppliers, clients, government agencies, and competitors during the adoption process. Regarding the use stage, subsection ``ICT use in MSEs'' showed that use is also mediated by external factors like the existence of specific ICT infrastructure (e.g., landlines), social inclusion (e.g., self-perceived social status), and cultural barriers (e.g., trust). Finally, the analysis presented in the subsection ``ICT impact on MSEs'' showed that impact assessments at the individual firm level could have severe limitations in understanding broader changes associated with value chains, informality, or market structures. In short, ICT and MSEs are embedded in broader social and institutional settings, and digital innovation processes can only be understood through conceptual frameworks capable of incorporating the complexity of these settings. 

Such a systemic perspective must also incorporate time. \textcite{Molla2006} show that such conceptual frameworks are vital in understanding change and evolution (both of ICTs and MSEs). This was supported by the literature highlighting the transitions from personal to business uses of mobile phones presented in the subsection ``ICT use in MSEs''.  Furthermore, since ICTs, MSEs, and their social and institutional context are dynamic, it makes sense to use such conceptual frameworks. It is also important to remember that using technologies is not a simple process as it implies co-evolution between the technology and the user \autocite{Fleck1993} and includes strong feedback loops to the adoption context \autocite{Geels2005}. This means that theoretical frameworks used in future research should be able to model and understand the change in the ICT, the MSE, the relationship between them, and the MSE's context. 

The final main conclusion, generalisability, is crucial because it has profound implications in selecting research and theoretical positions adopted in future research and the policy and practice implications of the field. As already discussed, digital innovation processes in SMEs are highly contingent in all its phases. The main implication of this contingency is that knowledge on this subject can seldom be generalised or extended to other cases. The results from this review indicate that the field should move towards analytical generalisability, in which research results are generalised to theoretical elements that could be used as analytical lenses to understand comparable realities \autocite{Walsham1995}. This does not mean that future research in this field cannot draw practical lessons that practitioners could use. However, these lessons should be left to the practitioners who are in a better position to judge what applies to their situation \autocite{Kennedy1979}. 

A final discussion of this review concerns some missing elements such as research on adoption side effects, systemic impact, and the provision of appropriate ICT for MSEs. The side effects refer to instances in which the adoption of an ICT (by individual MSEs, their market, or their value chain) leads to negative impacts on the MSE. Systemic impacts refer to the systemic outcomes of digital innovation processes like market structure reconfiguration, power shifts, and emergence of new institutions, among others. Finally, the provision of appropriate ICT is related to the issues of cost and short-term mentality as adoption barriers presented in the subsection ``ICT adoption in MSEs''. Although the literature identifies these adoption barriers well, there is a notable absence of work questioning the role of ICT producers in the innovation processes, as ICTs were usually black-boxed and presented as static.  
\section*{Statements and Declarations}
\label{sec:org724acf4}
No funding was received for conducting this study. The authors have no relevant financial or non-financial interests to disclose. 

\printbibliography
\end{document}